\documentclass{PoS}

\title{Dijet correlations in $pp$ collisions at RHIC}

\ShortTitle{Dijet correlations in $pp$ collisions at RHIC}

\author{\speaker{Antoni Szczurek}\\
        Institute of Nuclear Physics, PL-31-342 Cracow, Poland\\
        and University of Rzesz\'ow, PL-35-959 Rzesz\'ow, Poland\\
        E-mail: \email{Antoni.Szczurek@ifj.edu.pl}}

\author{Anna Rybarska\\
        Institute of Nuclear Physics, PL-31-342 Cracow, Poland\\
        E-mail: \email{Anna.Rybarska@ifj.edu.pl}}

\author{Gabriela Slipek\\
        Institute of Nuclear Physics, PL-31-342 Cracow, Poland\\
        E-mail: \email{Gabriela.Slipek@ifj.edu.pl}}

\abstract{
We compare results of $k_t$-factorization approach and next-to-leading
order collinear-factorization approach for dijet correlations in
proton-proton collisions at RHIC energies.
We discuss correlations in azimuthal angle as well as 
correlations in two-dimensional space of transverse momenta of two jets.
Some $k_t$-factorization subprocesses are included for the first
time in the literature.
Different unintegrated gluon/parton distributions are used in
the $k_t$-factorization approach. The results depend on UGDF/UPDF used.
Limitations due to leading jet condition are discussed.
}

\FullConference{High-pT physics at LHC\\
		 March 23-27 2007\\
		 University of Jyv\"askyl\"a, Jyv\"askyl\"a, Finland}

\begin{document}
\section{Introduction}

The jet correlations are interesting in the context
of recent detailed studies of hadron-hadron correlations
in nucleus-nucleus \cite{RHIC_correlations_nucleus_nucleus}
and proton-proton \cite{RHIC_correlations_proton_proton} collisions.
Those studies provide interesting information on the dynamics of
nuclear and elementary collisions.
Effects of geometrical jet structure were discussed recently
in Ref.\cite{Levai}. No QCD calculation of parton radiation was performed
up to now in this context. Before going into hadron-hadron correlations it
seems indispensable to better understand correlations between jets
due to the QCD radiation.
In this paper we address the case of elementary hadronic collisions in order
to avoid complicated and not yet well understood nuclear effects.
Our analysis should be considered as a first step in order to understand
the nuclear case in the future. 
In leading-order collinear-factorization approach jets are produced
back-to-back. These leading-order jets are therefore not included into
correlation function, although they contribute a big ($\sim
\frac{1}{2}$) fraction to the inclusive cross section.

\begin{figure}[!h]    
\begin{center}
 {\includegraphics[width=0.3\textwidth]{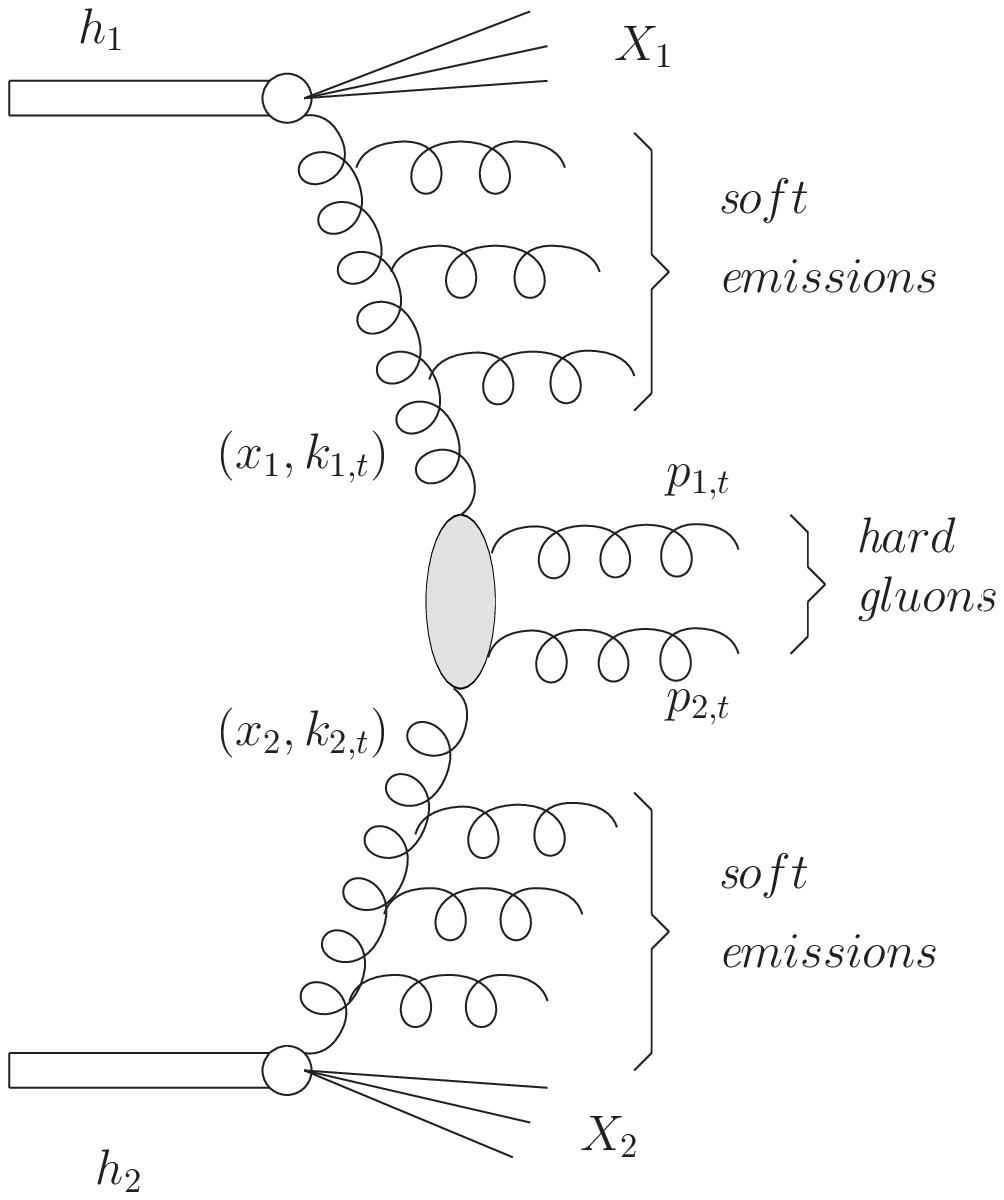}}
 {\includegraphics[width=0.3\textwidth]{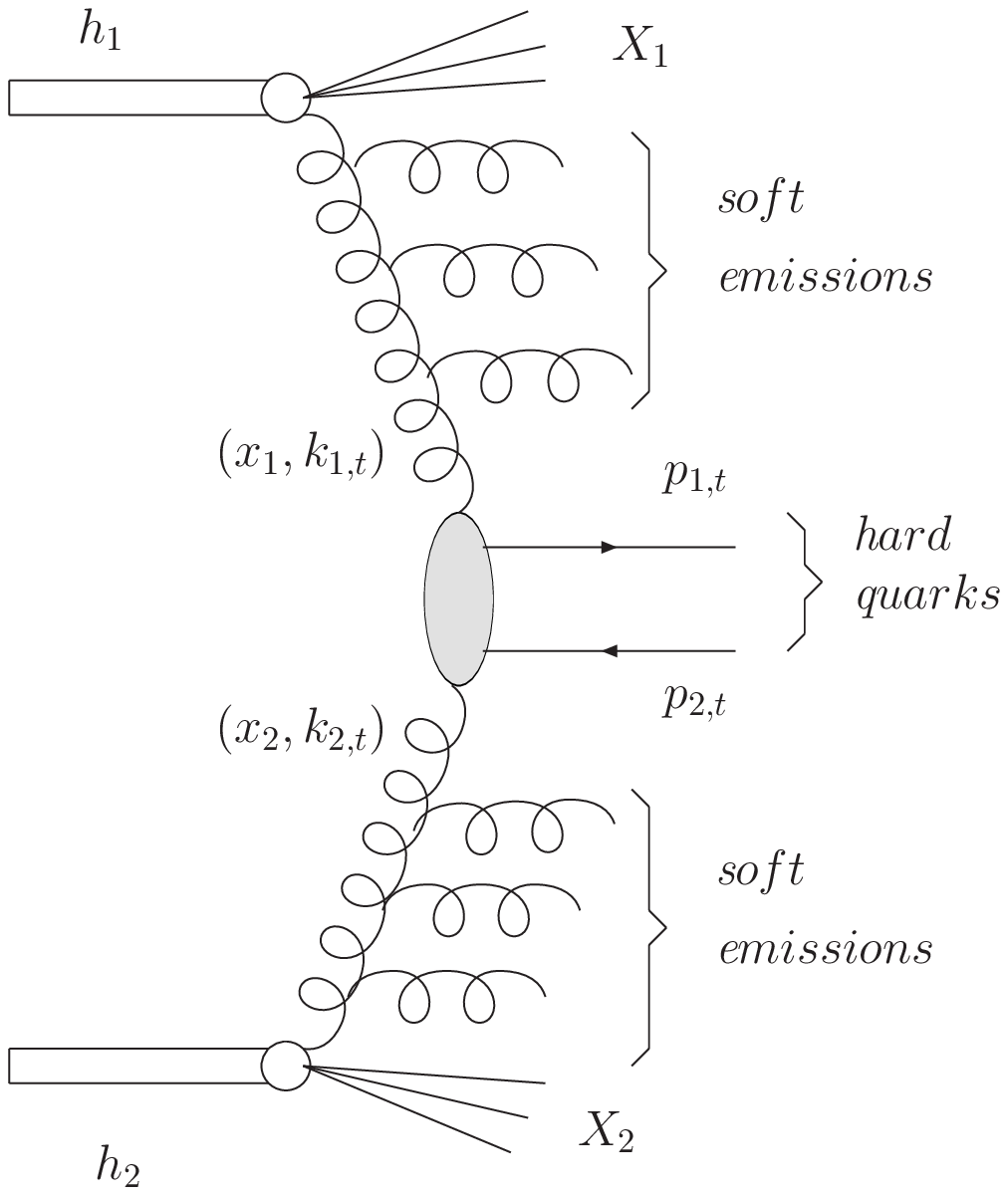}}
\end{center}
   \caption{\label{fig:kt_factorization_diagrams_old}
   \small  Diagrams for $k_t$-factorization approach included in the
literature. We shall call them $A_1$ and $A_2$ for brevity.}
\end{figure}

The truly internal momentum distribution
of partons in hadrons due to Fermi motion (usually neglected in
the literature) and/or any soft emission would lead to a decorrelation
from the simple kinematical back-to-back configuration.
In the fixed-order collinear approach only next-to-leading order terms
lead to nonvanishing cross sections at $\phi \ne \pi$ and/or
$p_{1,t} \ne p_{2,t}$ (moduli of transverse momenta of outgoing partons).
In the $k_t$-factorization approach, where transverse momenta
of gluons entering the hard process are included explicitly,
the decorrelations come naturally in a relatively easy to calculate way.
In Fig.\ref{fig:kt_factorization_diagrams_old} we show
$k_t$-factorization processes discussed up to now in the literature
\cite{LO00,O00,Bartels}.
The soft emissions, not explicit
in our calculation, are hidden in model unintegrated gluon distribution
functions (UGDF). In our calculation the last objects are assumed to be
given and are taken from the literature
\cite{KL01,AKMS94,kwiecinski}.


\begin{figure}[!h]    
\begin{center}
 {\includegraphics[width=0.3\textwidth]{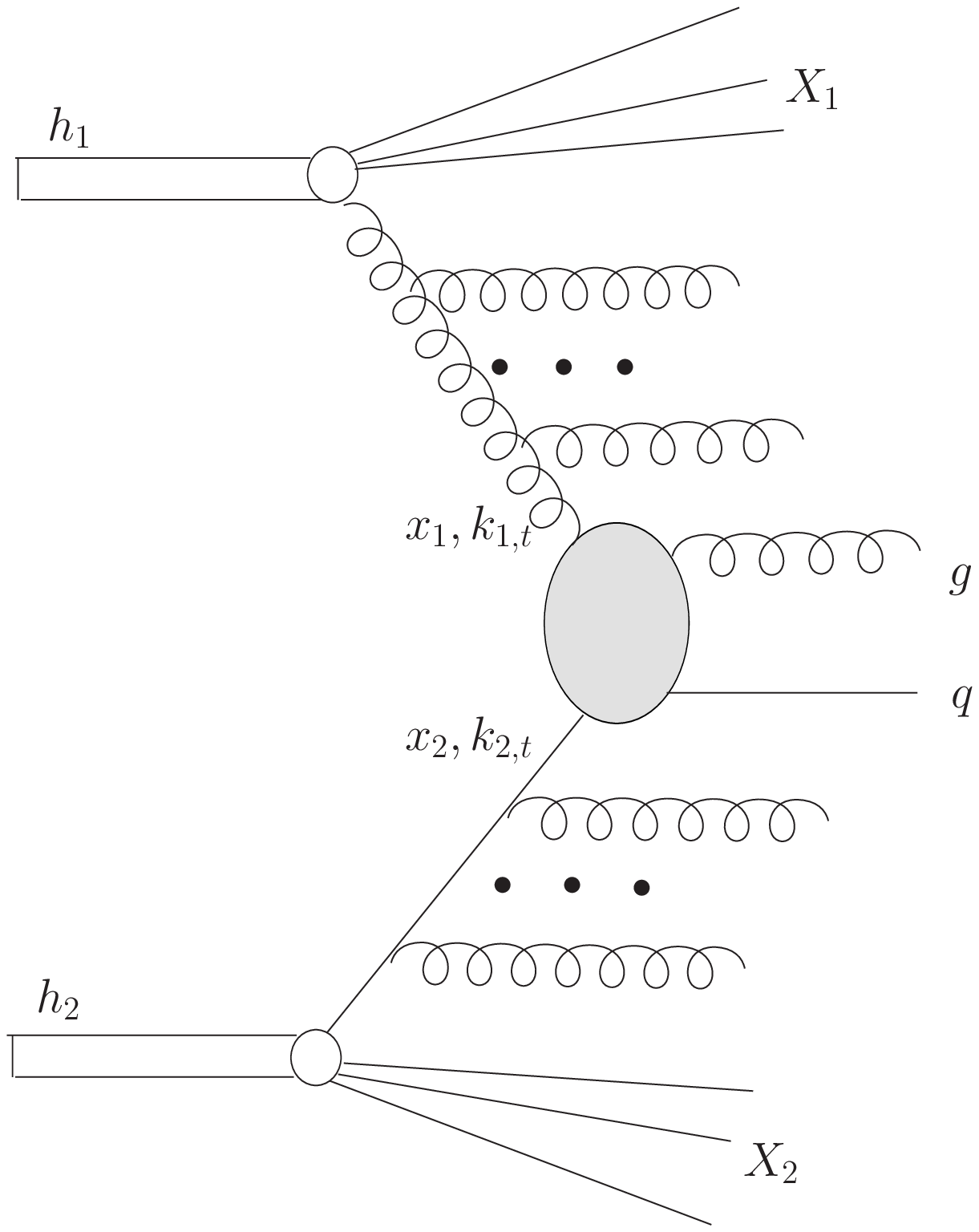}}
 {\includegraphics[width=0.3\textwidth]{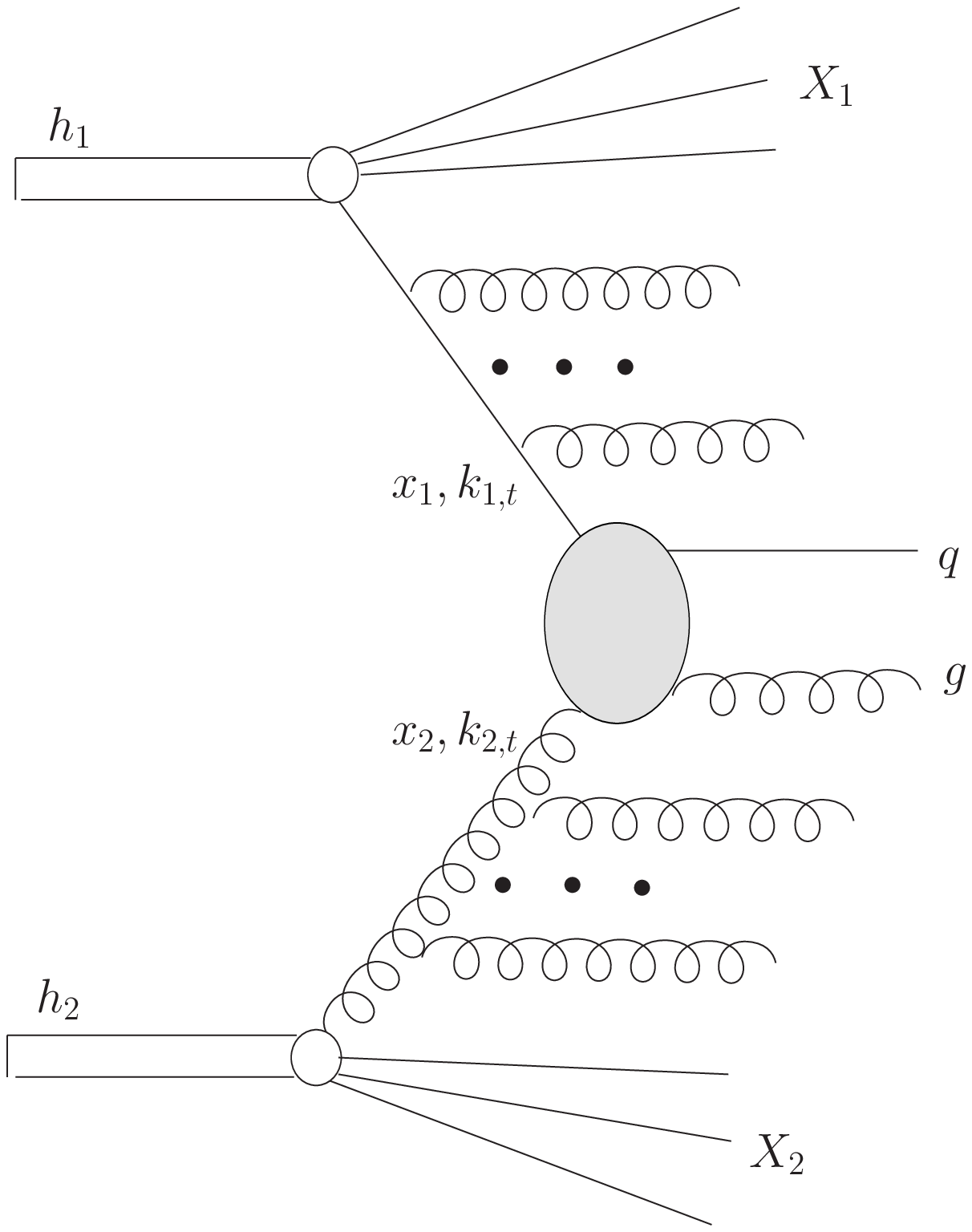}}
\end{center}
   \caption{\label{fig:kt_factorization_diagrams_new}
   \small  Diagrams for $k_t$-factorization approach included for the
first time here. We shall call them $B_1$ and $B_2$ for brevity.}
\end{figure}

In addition we include two new processes 
(see Fig.\ref{fig:kt_factorization_diagrams_new}), not discussed
up to now in the context of $k_t$-factorization approach. 
We shall discuss their role at the RHIC energy $W$ = 200 GeV.


Furthermore we compare results obtained within the $k_t$-factorization
approach and results obtained in the NLO collinear-factorization.
Here we wish to address the problem of the relation between both approaches.
We shall identify the regions of the phase space where the
hard $2 \to 3$ processes, not explicitly included in the
leading-order $k_t$-factorization approach, dominate over the
$2 \to 2$ contributions calculated with UGDFs. 
We shall show how the results depend on UGDFs used.  

We shall concentrate on the region of relatively semi-hard jets, i.e.
on the region related to the recently measured hadron-hadron
correlations at RHIC. 

\section{Formalism}

The cross section for the production of a pair of partons (k,l)
can be written as
\begin{eqnarray}
\frac{d\sigma(h_1 h_2 \rightarrow jet jet)}
{d^2p_{1,t}d^2p_{2,t}} &=& \sum_{i,j,k,l} \int dy_1 dy_2
\frac{d^2 k_{1,t}}{\pi}\frac{d^2 k_{2,t}}{\pi}
\frac{1}{16\pi^2(x_1x_2s)^2}
\overline{|{\cal M}(i j \rightarrow k l)|^2}
\nonumber \\
&\cdot&\delta^2(\overrightarrow{k}_{1,t}
+\overrightarrow{k}_{2,t}
-\overrightarrow{p}_{1,t}
-\overrightarrow{p}_{2,t})
{\cal F}_i(x_1,k_{1,t}^2){\cal F}_j(x_2,k_{2,t}^2) \; ,
\label{basic_formula}
\end{eqnarray}
where
\begin{equation}
x_1 = \frac{m_{1,t}}{\sqrt{s}}\mathrm{e}^{+y_1} 
    + \frac{m_{2,t}}{\sqrt{s}}\mathrm{e}^{+y_2} \; ,
\end{equation}
\begin{equation}
x_2 = \frac{m_{1,t}}{\sqrt{s}}\mathrm{e}^{-y_1} 
    + \frac{m_{2,t}}{\sqrt{s}}\mathrm{e}^{-y_2} \; ,
\end{equation}
and $m_{1,t}$ and $m_{2,t}$ are so-called transverse masses
defined as $m_{i,t} = \sqrt{p_{i,t}^2+m^2}$, where $m$ is the mass
of a parton. In the following we shall assume that all partons are
massless.
The objects denoted by ${\cal F}_i(x_1,k_{1,t}^2)$ and
${\cal F}_j(x_2,k_{2,t}^2)$ in the equation (\ref{basic_formula}) above
are the unintegrated parton distributions in hadron $h_1$ and $h_2$,
respectively.
They are functions of longitudinal momentum fraction and transverse
momentum of the incoming (virtual) parton.

After some simple algebra one obtains a handy formula:
\begin{eqnarray}
\frac{d \sigma(p_{1,t},p_{2,t})}{d p_{1,t} d p_{2,t}}
&&= \frac{1}{2} \cdot \frac{1}{2} \cdot 4 \pi
\int d \phi_{-} \; p_{1,t} p_{2,t} \;
\sum_{i,j,k,l} \int d y_1 d y_2 \; \frac{1}{4} q_t d q_t d \phi_{q_t}
\nonumber \\
&& \left(
\frac{1}{16\pi^2(x_1x_2s)^2}
\overline{|{\cal M}(i j \rightarrow k l)|^2}
{\cal F}_i(x_1,k_{1,t}^2){\cal F}_j(x_2,k_{2,t}^2)
\right) \; .
\label{2-dim-map_c}
\end{eqnarray}
This 5-dimensional integral is now calculated for each point on the map
$p_{1,t} \times p_{2,t}$.

\begin{figure}[!h]    
\begin{center}
 \centerline{\includegraphics[width=0.5\textwidth]{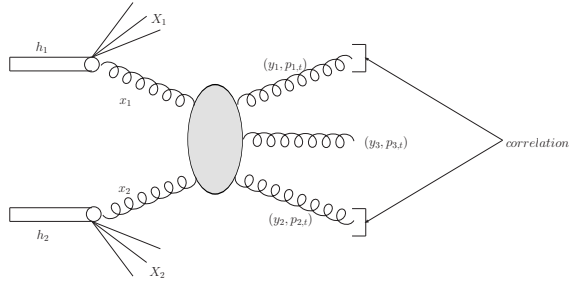}}
\end{center}
   \caption{\label{fig:NLO_diagrams}
   \small  A typical $2 \to 3$ process.
The kinematical variables used are shown explicitly.}
\end{figure}

Up to now we have considered only processes with two explicit hard
partons. 
Now we shall discuss also processes with three explicit hard partons.
In Fig.\ref{fig:NLO_diagrams} we show a typical $2 \to 3$ process
and kinematical variables needed in the description of
the process. We select the particle 1 and 2 as those which correlations
are studied. This is only formal as all possible combinations are
considered in real calculations.


The cross section for $h_1 h_2 \to g g g X$ can be calculated according
to the standard parton model formula:
\begin{equation}
d \sigma (h_1 h_2 \to g g g) =
\int d x_1 d x_2 \; g_1(x_1,\mu^2) g_2(x_2,\mu^2) \; 
d {\hat \sigma}(g g \to g g g)  \; ,
\label{2to3_parton_formula}
\end{equation}
where the longitudinal momentum fractions are evaluated as
\begin{eqnarray}
x_1 &=& \frac{p_{1,t}}{\sqrt{s}} \exp(+y_1)
      + \frac{p_{2,t}}{\sqrt{s}} \exp(+y_2)
      + \frac{p_{3,t}}{\sqrt{s}} \exp(+y_3) \; ,\nonumber \\
x_2 &=& \frac{p_{1,t}}{\sqrt{s}} \exp(-y_1)
      + \frac{p_{2,t}}{\sqrt{s}} \exp(-y_2)
      + \frac{p_{3,t}}{\sqrt{s}} \exp(-y_3) \; .
\end{eqnarray}
After a simple algebra \cite{SRS07} we get finally:
\begin{equation}
d \sigma = \frac{1}{64 \pi^4 \hat{s}^2} \; 
x_1 g_1(x_1,\mu_f^2) x_2 g_2(x_2,\mu_f^2) \; \overline{|{\cal M}_{2 \to 3}|^2}
p_{1,t} dp_{1,t} p_{2,t} dp_{2,t} d \phi_{-} dy_1 dy_2 dy_3 \; ,
\label{2to3_handy_formula}
\end{equation}
where $\phi_{-}$ is restricted to the interval $(0,\pi)$.
The last formula is very useful in calculating the cross section
for particle 1 and particle 2 correlations.

\section{Results}

Let us concentrate first on $2 \to 2$ processes calculated within
$k_t$-factorization approach.

\begin{figure}[!h]      
\begin{center}
{\includegraphics[width=0.35\textwidth]{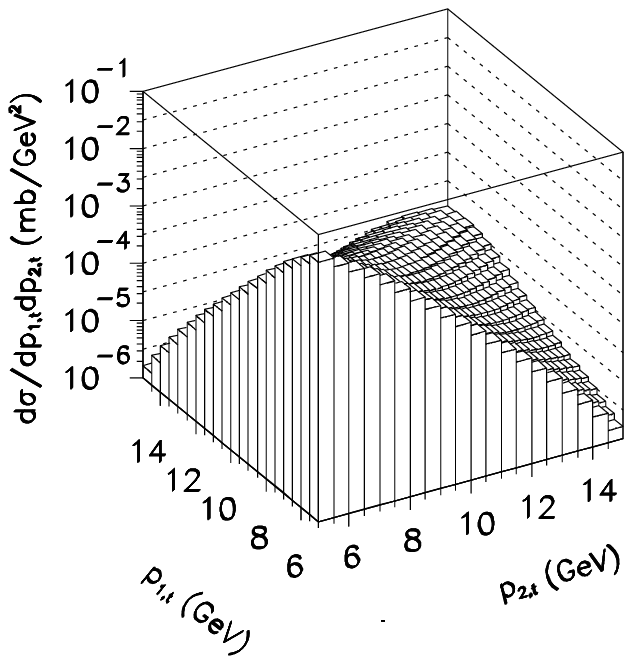}}
{\includegraphics[width=0.35\textwidth]{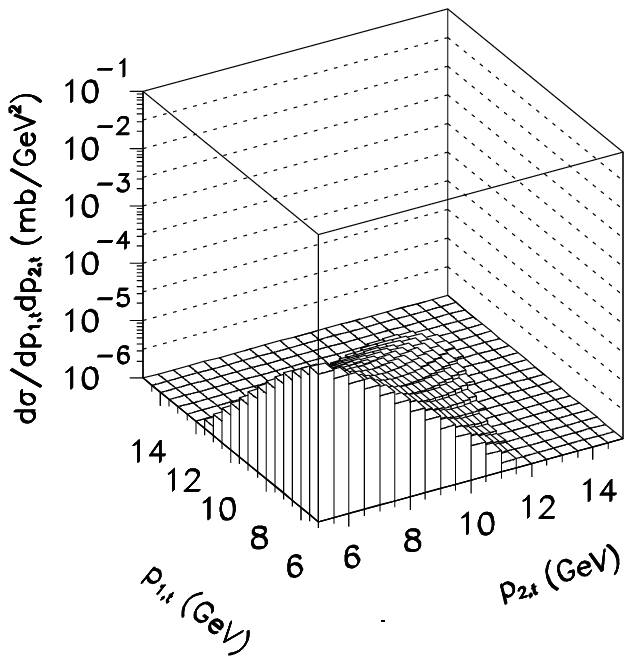}}
\\
{\includegraphics[width=0.35\textwidth]{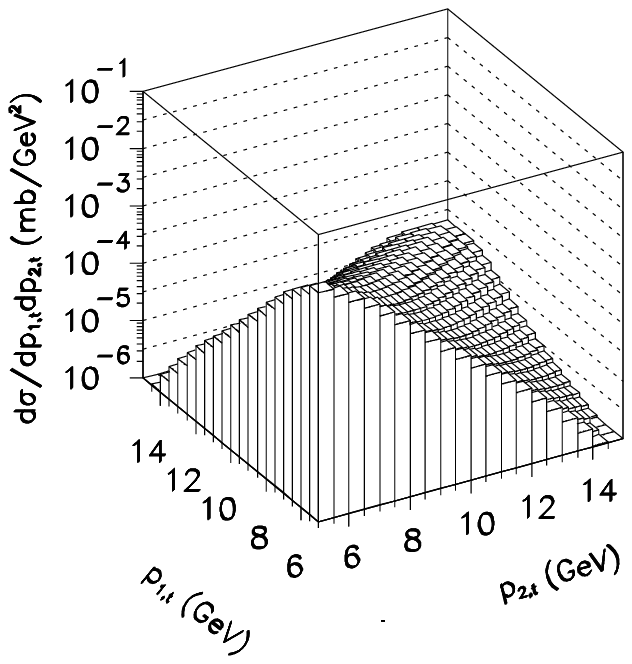}}
{\includegraphics[width=0.35\textwidth]{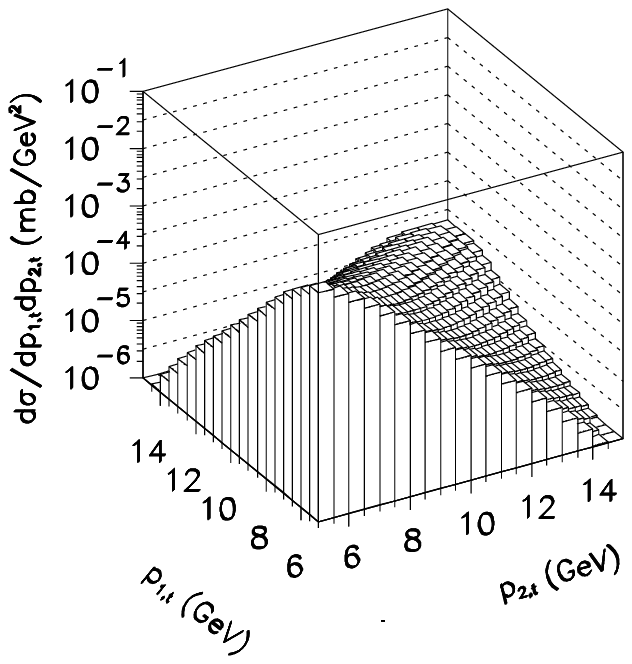}}
\end{center}
\caption{\small
Two-dimensional distributions in $p_{1,t}$ and $p_{2,t}$ 
for different subprocesses $gg \to gg$ (left upper)
$gg \to q \bar q$ (right upper), $gq \to gq$ (left lower)
and $qg \to qg$ (right lower). In this calculation W = 200 GeV and
Kwieci\'nski UPDFs with exponential nonperturbative form factor
($b_0$ = 1 GeV$^{-1}$) and $\mu^2$ = 100 GeV$^2$ were used.
Here integration over full range of parton rapidities was made.
\label{fig:p1tp2t_kwiec_components}
}
\end{figure}


In Fig.\ref{fig:p1tp2t_kwiec_components} we show two-dimensional
maps in $(p_{1,t},p_{2,t})$ for all $k_t$-factorization processes
shown in Fig.\ref{fig:kt_factorization_diagrams_old} and 
Fig.\ref{fig:kt_factorization_diagrams_new}.
Only very few approaches in the literature include both gluons and
quarks and antiquarks.
In the calculation above we have used Kwieci\'nski UPDFs with
exponential nonperturbative form factor ($b_0$ = 1 GeV$^{-1}$)
and the factorization scale $\mu^2 = (p_{t,min}+p_{t,max})^2/4$ = 100 GeV$^2$.

\begin{figure}[!h]
\begin{center} 
{\includegraphics[width=0.35\textwidth]{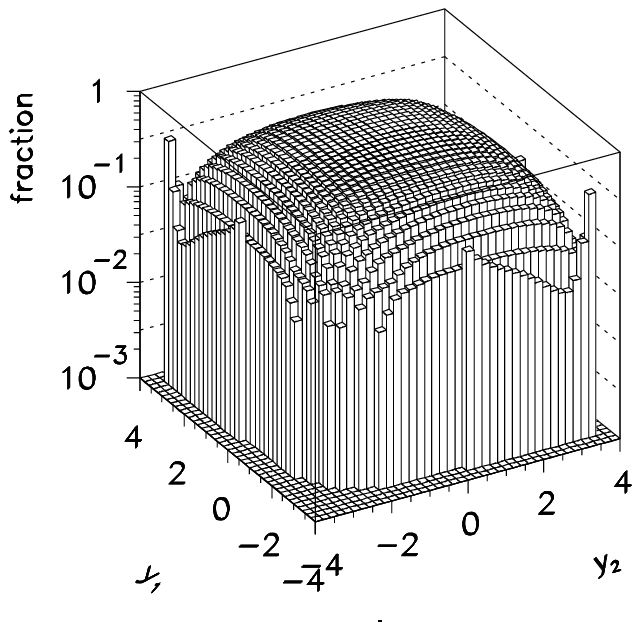}}
{\includegraphics[width=0.35\textwidth]{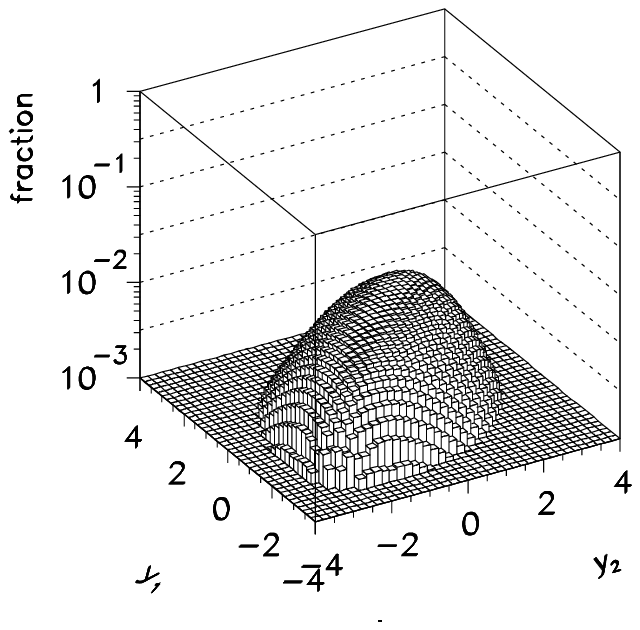}} \\
{\includegraphics[width=0.35\textwidth]{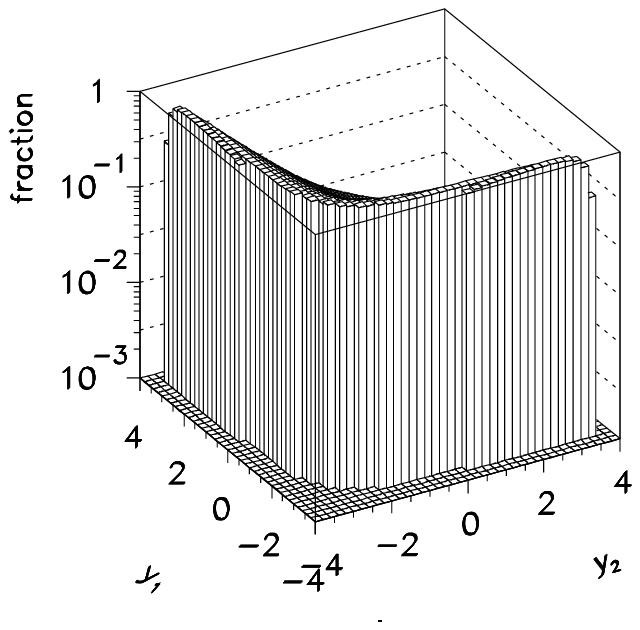}}
{\includegraphics[width=0.35\textwidth]{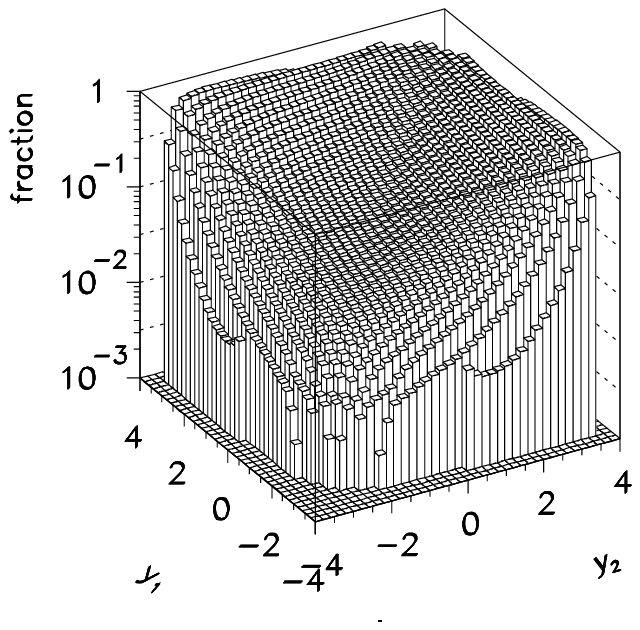}}
\end{center}     
\caption{\small
Two-dimensional distributions of fractional contributions of different
subprocesses
as a function of $y_1$ and $y_{2}$ 
for $gg \to gg$ (left upper)
$gg \to q \bar q$ (right upper), $gq \to gq$ (left lower)
and $qg \to qg$ (right lower). In this calculation W = 200 GeV and
Kwieci\'nski UPDFs with exponential nonperturbative form factor
and $b_0$ = 1 GeV$^{-1}$ were used. The integration is made
for jets from the transverse momentum interval:
5 GeV $< p_{1,t}, p_{2,t} <$ 20 GeV.
\label{fig:2to2_contributions_y1y2}}
\end{figure}


In Fig.\ref{fig:2to2_contributions_y1y2} we show fractional
contributions (individual component to the sum of all four components)
of the above four processes on the two-dimensional map $(y_1,y_2)$.
One point here requires a better clarification.
Experimentally it is not possible to distinguish gluon and
quark/antiquark jets. Therefore in our calculation of the $(y_1,y_2)$
dependence one has to symmetrize the cross section (not the amplitude)
with respect to gluon -- quark/antiquark exchange 
($y_1 \to y_2, y_2 \to y_1$).
While at midrapidities the contribution of diagram $B_1$ + $B_2$
is comparable to the diagram $A_1$, at larger rapidities the
contributions of diagrams of the type B dominate. 
The contribution of diagram $A_2$ is relatively
small in the whole phase space. When calculating the contributions of
the diagram $A_1$ and $A_2$ one has to be careful about collinear
singularity which leads to a significant enhancement of the cross section
at $\phi_{-}$=0 and $y_1 = y_2$, i.e. in the one jet case, when both
partons are emitted in the same direction.
This is particularly important for the matrix elements obtained by
the naive analytic continuation from the formula for on-shell initial partons.
The effect can be, however, easily eliminated with the jet-cone separation
algorithm \cite{SRS07}.

\begin{figure}[!h]   
 \centerline{\includegraphics[width=0.40\textwidth]{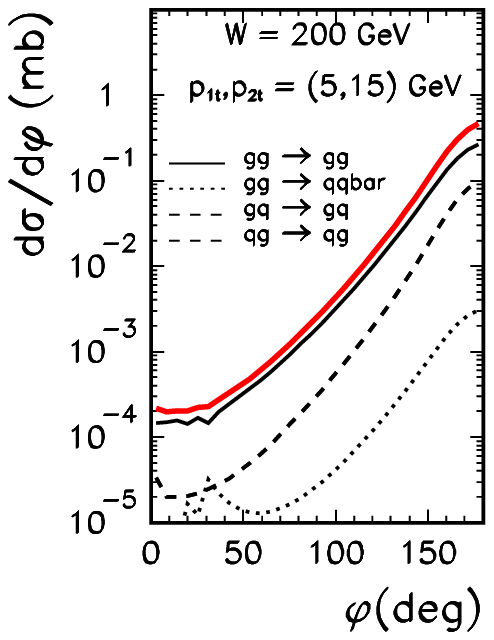}}
   \caption{ \label{fig:2to2_contributions_phi} 
\small 
The angular correlations for all four components: $gg \to gg$ (solid),
$gg \to q \bar q$ (dashed) and $gq \to gq$ = $qg \to qg$ (dash-dotted).
The calculation is performed with the Kwieci\'nski UPDFs and
$b_0$ = 1 GeV$^{-1}$.
The integration is made for jets from the transverse momentum interval:
5 GeV $< p_{1,t}, p_{2,t} <$ 15 GeV and from the rapidity interval:
-4 $< y_1, y_2 <$ 4.
}
\end{figure}


For completeness in Fig.\ref{fig:2to2_contributions_phi} we show
azimuthal angle dependence of the cross section for all four components.
There is no sizeable difference in the shape of azimuthal distribution
for different components.

\begin{figure}[!h]   
 \centerline{\includegraphics[width=0.4\textwidth]{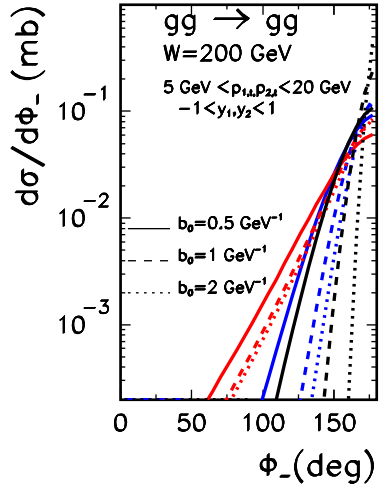}}
   \caption{ \label{fig:b0_mu2_phi}
\small
The azimuthal correlations for the $gg \to gg$ component obtained with
the Kwieci\'nski UGDFs for different values of the nonperturbative
parameter $b_0$ and for different evolution scales $\mu^2$ = 10 (on line
blue), 100 (on line red) GeV$^2$.
The initial distributions (without evolution) are shown for reference
by black lines.
}
\end{figure}

The Kwieci\'nski approach allows to separate the unknown perturbative
effects incorporated via nonperturbative form factors
and the genuine effects of QCD evolution.
The Kwieci\'nski distributions have two external parameters:
\begin{itemize}
\item the parameter $b_0$ responsible for nonperturbative effect
     (for details see \cite{SRS07}),
\item the evolution scale $\mu^2$ (for details see \cite{SRS07}).
\end{itemize}
While the latter can be identified physically with characteristic
kinematical quantities in the process $\mu^2 \sim p_{1,t}^2, p_{2,t}^2$,
the first one is of nonperturbative origin and cannot be calculated
from first principles.
The shapes of distributions depends, however, strongly on the value of
the parameter $b_0$ in which the inital momentum distribution is encoded.
This is demonstrated in Fig.\ref{fig:b0_mu2_phi} where we show angular
correlations in azimuth for the $gg \to gg$ subprocess.
The smaller $b_0$ the bigger decorrelation in azimuthal angle
can be observed. In Fig.\ref{fig:b0_mu2_phi} we show also the role of
the evolution scale in the Kwieci\'nski distributions.
The QCD evolution embedded in the Kwieci\'nski evolution equations
populate larger transverse momenta of partons entering the hard process.
This significantly increases the initial (nonperturbative) decorrelation
in azimuth.
For transverse momenta of the order of $\sim$ 10 GeV the effect of
evolution is of the same order of magnitude as the effect due to
the nonperturbative physics of hadron confinement. For larger scales
of the order of $\mu^2 \sim$ 100 GeV$^2$, more adequate for jet
production, the initial condition is of minor importance and the effect
of decorrelation is dominated by the evolution. Asymptotically (infinite
scales) there is no dependence on the initial condition provided
reasonable initial conditions are taken.

\begin{figure}[!h]
\begin{center}      
{\includegraphics[width=0.35\textwidth]{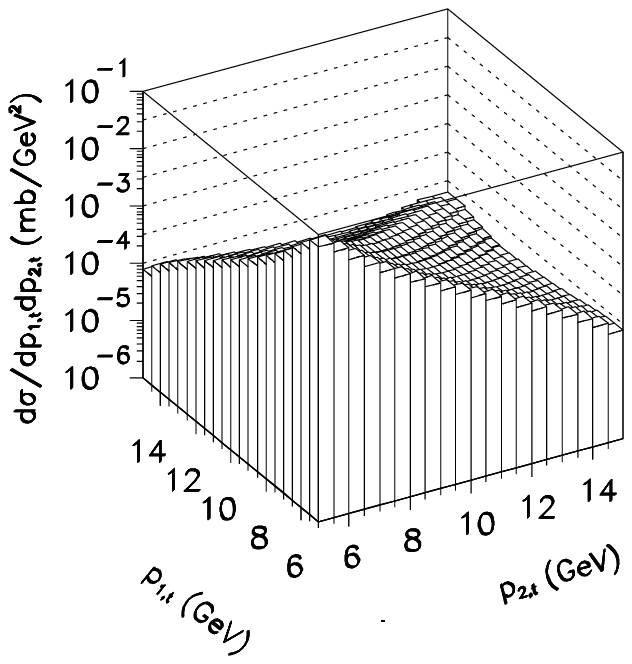}}
{\includegraphics[width=0.35\textwidth]{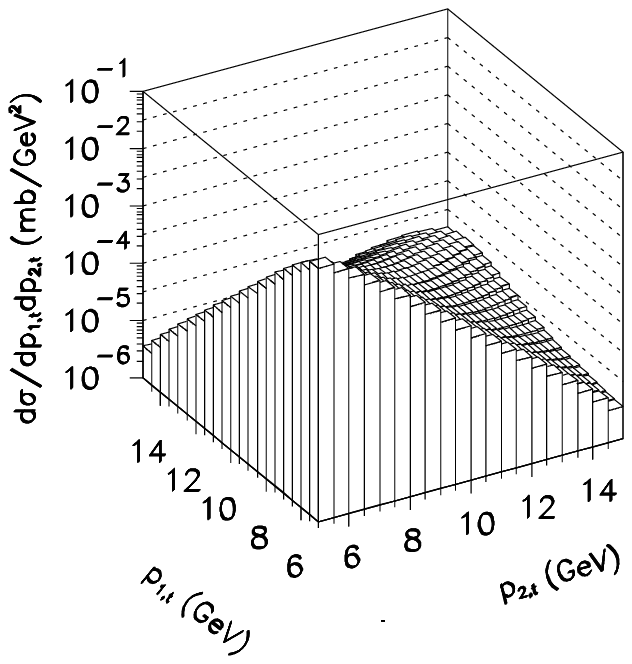}} \\
{\includegraphics[width=0.35\textwidth]{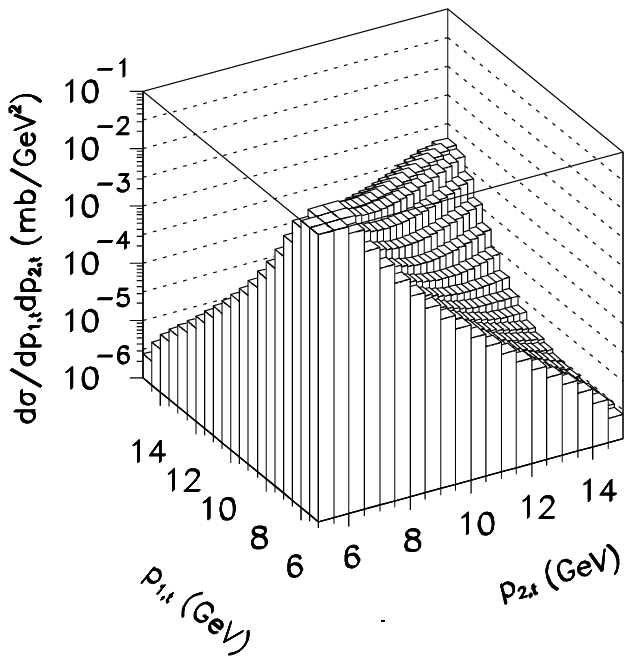}}
{\includegraphics[width=0.35\textwidth]{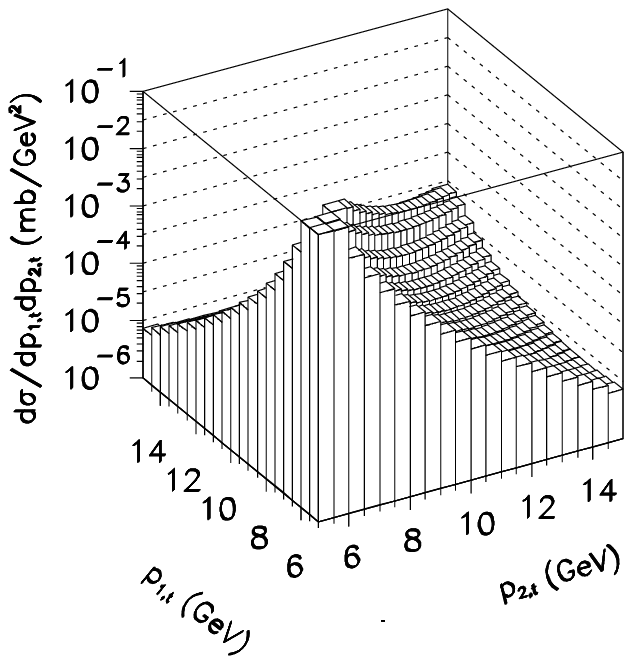}}     
\end{center}
\caption{\small
Two-dimensional distributions in $p_{1t}$ and $p_{2t}$ for
KL (left upper), BFKL (right upper), Ivanov-Nikolaev (left lower) UGDFs
and for the $g g \to g g g$ (right lower). In this calculation
-4 $< y_1, y_2<$ 4.
\label{fig:p1tp2t_maps}}
\end{figure}


In Fig.\ref{fig:p1tp2t_maps} we show the maps for different
UGDFs and for $gg \to ggg$ processes in the broad range of transverse
momenta 5 GeV $< p_{1,t},p_{2,t} <$ 20 GeV for the RHIC energy W = 200
GeV. In this calculation we have not imposed any particular cuts
on rapidities. We have not imposed also any cut on the transverse momentum
of the unobserved third jet in the case of $2 \to 3$ calculation.
The small transverse momenta of the third jet contribute to the sharp
ridge along the diagonal $p_{1,t} = p_{2,t}$.
Naturally this is therefore very difficult to distinguish these three-parton
states from standard two jet events.
In principle, the ridge can be eliminated by imposing a cut on the
transverse momentum of the third (unobserved) parton \cite{SRS07}.
There are also other methods to eliminate the ridge and underlying soft
processes which is discussed in Ref.\cite{SRS07}.

\begin{figure}[!h]   
 \centerline{\includegraphics[width=0.40\textwidth]{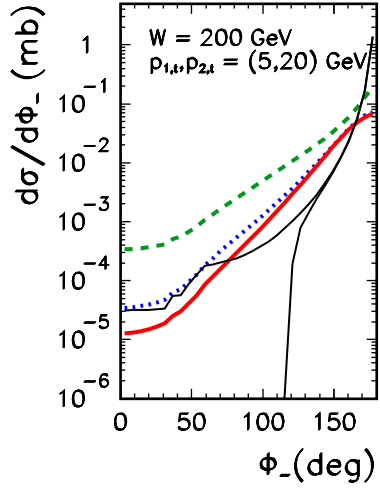}}
   \caption{ \label{fig:dsig_dphid_updf_vs_ljNLO}
\small 
Dijet azimuthal correlations $d \sigma / d\phi_{-}$
for the $gg \to gg$ component and different UGDFs
as a function of azimuthal angle between the gluonic jets. 
In this calculation W = 200 GeV and -1 $< y_1, y_2<$ 1,
5 GeV $< p_{1t}, p_{2t} <$ 20 GeV. Here the thick-solid line corresponds
to the Kwieci\'nski UGDF, the dashed line to the Kharzeev-Levin type
of distribution and the dotted line to the BFKL distribution.
The two thin solid (on line black) lines
are for NLO collinear approach without (upper line) and with (lower
line) leading jet restriction. 
}
\end{figure}


\begin{figure}[!h]
 \centerline{\includegraphics[width=0.40\textwidth]{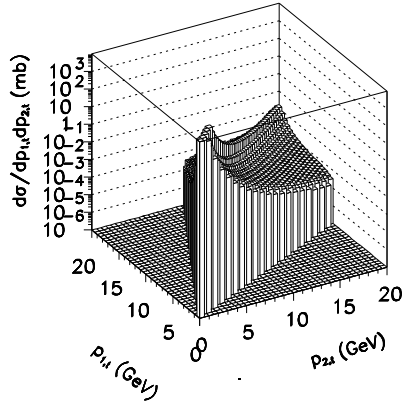}}
   \caption{ \label{fig:map_p1tp2t_lj}
\small
Cross section for the $gg \to ggg$ component on the $(p_{1,t},p_{2,t})$ plane
with the condition of leading jets (partons).
}
\end{figure}


When calculating dijet correlations
in the standard NLO ($2 \to 3$) approach we have taken all possible dijet
combinations. This is different from what is usually taken in
experiments \cite{D0_data}, where correlation between leading jets
are studied. In our notation this means $p_{3,t} < p_{1,t}$ and
$p_{3,t} < p_{2,t}$. When imposing such extra condition on our NLO
calculation we get the dash-dotted curve in
Fig.\ref{fig:dsig_dphid_updf_vs_ljNLO}.
In this case $d \sigma / d \phi_{-} = 0$ for $\phi_{-} < 2/3 \pi$.
This vanishing of the cross section is of purely kinematical origin.
Since in the $k_t$-factorization calculation only two jets are explicit,
there is no such an effect in this case.
This means that the region of $\phi_{-} < 2/3 \pi$ should
be useful to test models of UGDFs.
For completeness in Fig.\ref{fig:map_p1tp2t_lj} we show a
two-dimensional plot $(p_{1,t},p_{2,t})$ with imposing the leading-jet
condition.
Surprisingly the leading-jet condition removes a big part of
the two-dimensional space. In particular, regions with $p_{2,t} > 2 p_{1,t}$ 
and $p_{1,t} > 2 p_{2,t}$ cannot be populated via
$2 \to 3$ subprocess \footnote{In LO collinear approach the whole plane,
except of the diagonal $p_{1,t} = p_{2,t}$, is forbidden.}.
There are no such limitations for $2 \to 4$, $2 \to 5$ and even
higher-order processes.
Therefore measurements in ``NLO-forbidden'' regions of the
$(p_{1,t},p_{2,t})$ plane
would test higher-order terms of the standard collinear pQCD.
These are also regions where UGDFs can be tested, provided that not too
big transverse momenta of jets are taken into the correlation in order
to assure the dominance of gluon-initiated processes. For larger
transverse momenta and/or forward/backward rapidities one has to include
also quark/antiquark initiated processes via unintegrated
quark/antiquark distributions.

\section{Summary}
Motivated by the recent experimental results of hadron-hadron correlations
at RHIC we have discussed dijet correlations in proton-proton collisions.
We have considered and compared results obtained
with collinear next-to-leading order approach and leading-order
$k_t$-factorization approach. 

In comparison to recent works in the framework of $k_t$-factorization
approach, we have included two new mechanisms based on $gq \to gq$
and $qg \to qg$ hard subprocesses.
This was done based on the Kwieci\'nski unintegrated parton
distributions.
We find that the new terms give significant contribution at RHIC energies.
In general, the results of the $k_t$-factorization approach depend
on UGDFs/UPDFs used, i.e. on approximation and assumptions made
in their derivation.

The results obtained in the standard NLO approach depend significantly
whether we consider correlations of any jets or correlations of only
leading jets. 
In the NLO approach one obtains
$\frac{d \sigma}{d \phi_{-}}$ = 0 if $\phi_{-} < 2/3 \pi$
for leading jets as a result of a kinematical constraint.
Similarly $\frac{d \sigma}{d p_{1,t} d p_{2,t}}$ = 0 if $p_{1,t} > 2
p_{2,t}$ or $p_{2,t} > 2 p_{1,t}$.

There is no such a constraint in the $k_t$-factorization approach
which gives a nonvanishing cross section at small relative azimuthal
angles between leading jets and transverse-momentum asymmetric
configurations. We conclude that in these regions the $k_t$-factorization
approach is a good and efficient tool for the description of leading-jet
correlations.
Rather different results are obtained with different UGDFs
which opens a possibility to verify them experimentally.
Alternatively, the NLO-forbidden configurations can be described
only by higher-order (NNLO and higher-order) terms.
We do not need to mention that this is a rather difficult and technically
involved computation.

What are consequences for particle-particle correlations measured
recently at RHIC requires a separate dedicated analysis.
Here the so-called leading particles may come both from leading
and non-leading jets.
This requires taking into account the jet fragmentation process.

\vspace{0.5cm}

{\bf Acknowledgments} This work is partially supported by the grant of the Polish Ministry
of Scientific Research and Information Technology number 1 P03B 028 28. 


\end{document}